# Effect of coordinate rotation on 3D molecular descriptors computed by DragonX

*M. Hechinger and W. Marquardt\**

[a]AVT-Process Systems Engineering, RWTH Aachen University, 52064 Aachen, Germany;

*Corresponding author's e-mail address: wolfgang.marquardt@avt.rwth-aachen.de; tel.: +49-241-80-94668; fax: +49-241-80-92326

## Abstract

Quantitative structure-property relations (QSPR) employing descriptors derived from the 3D molecular structure are frequently applied for property prediction in various fields of research. In particular, DragonX is one of the most widely used software packages for descriptor calculation. The reliability of 3D molecular descriptors computed by DragonX has lately been investigated, thereby focusing on the effect of computational methods used for molecular structure optimization on the accuracy of the resulting molecular descriptors. The present contribution extends the analysis to a more intrinsic problem of DragonX descriptor evaluation resulting from the sensitivity of the computed 3D descriptors on the coordinate system used for molecule description. Evaluating several 3D descriptors for converged molecular structures rotated around all 3 spatial axes (affine coordinate transformations) yields systematically varying descriptor values. Since this unphysical behavior severely affects the descriptor reliability, the present contribution is meant to summarize these findings for later improvement of either descriptor definitions or the DragonX implementations.

## Computation of molecular structure

The molecular structures of *n*-propane, *n*-butane and 2-methylpentane have been computed as described in [1]. Hence, for each compound a systematic conformer search with the semi-empirical AM1 method [2] has been performed using Spartan10 [3]. Subsequently, each conformer has been optimized in Gaussian09 [4] using the B3LYP method with a TZVP basis set. Moreover, the partial atomic charges which are required for the evaluation of charge descriptors in DragonX [5] have been computed based on a B3LYP computation with an aug-cc-pVTZ basis set, since the additional diffuse basis functions allow for a better representation of the electrostatic behavior. The partial charges have then been obtained by fitting to the electrostatic potential according to the CHELP scheme [6] in Gaussian09 [4].

## Evaluation of 3D descriptors

Based on the molecular information obtained from the computations described in the previous section, 3D molecular descriptors have been determined using DragonX [5] for *n*-propane, *n*-butane and 2-methylpentane. DragonX computes a total of 735 molecular descriptors from 6 different classes which depend on the 3D molecular structure or atomic charge. To this end, molecular structures have been passed to DragonX in the Sybyl mol2-format [7], which contains both, the atomic coordinates $c_x$, $c_y$ and $c_z$ as well as the atomic partial charges.

For each molecule, the original coordinates

$$C = \begin{pmatrix} c_{1,x} & c_{1,y} & c_{1,z} \\ \vdots & \vdots & \vdots \\ c_{n,x} & c_{n,y} & c_{n,z} \end{pmatrix} \qquad (1)$$

are rotated around the *x*-, *y*- and *z*-axis by multiplication with the transformation matrices

$$R_x = \begin{pmatrix} 1 & 0 & 0 \\ 0 & cos(\phi) & -sin(\phi) \\ 0 & sin(\phi) & cos(\phi) \end{pmatrix}, \qquad (2)$$

$$R_y = \begin{pmatrix} \cos(\phi) & 0 & \sin(\phi) \\ 0 & 1 & 0 \\ -\sin(\phi) & 0 & \cos(\phi) \end{pmatrix}, \tag{3}$$

$$R_z = \begin{pmatrix} \cos(\phi) & -\sin(\phi) & 0 \\ \sin(\phi) & \cos(\phi) & 0 \\ 0 & 0 & 1 \end{pmatrix}, \tag{4}$$

respectively. In eqs. (1)-(4), $n$ denotes the number of atoms in the molecule and $\phi$ denotes the rotation angle. The transformed coordinates

$$C_i(\phi) = R_i(\phi) \cdot C, \quad i \in \{x, y, z\} \tag{5}$$

thus represent the same molecule $C$, however, with respect to the new coordinate system which rotated by the angle $\phi$ around the respective axis. Since the rotation is an affine translation, it leaves the distances between the atoms unchanged, such that molecular descriptors depending on interatomic distances should not be affected.

For each of the molecules *n*-propane, *n*-butane and 2-methylpentane, all 735 3D molecular descriptors have been computed for rotation angles between 0° and 360° in steps of 10°, thereby rotating each molecule around the *x*-, the *y*- and the *z*-axis subsequently. That is, while rotating around one axis, the angle for the other two axes has been fixed to 0°, such that rotations have only been performed around one axis at a time.

## Results and discussion

The systematic conformer search resulted in the number of conformations as shown in Table 1. Hence, a total of 10 different conformations has been investigated.

**Table 1:** Number of conformations identified by Spartan10 [3] for each of the compounds.

| Method | *n*-propane | *n*-butane | 2-methylpentane |
|---|---|---|---|
| # conformations | 1 | 2 | 7 |

Figure 1 shows the variation of the descriptor DISPv of *n*-propane as a function of the rotation angle $\phi$ around all three axes. As can been seen, for rotations of the original molecular structure around the *x*- and the *y*-axis there exists a systematic variation of the descriptor value with a maximum value of 0.65 at angles of $k \cdot \pi/4$ (for $k = 1, 2, 3, 4$) and a minimum value of 0 at $k \cdot \pi/2$ (for $k = 0, 1, 2, 3$). When rotating the molecule around the *z*-axis, however, the descriptor value equals 0 for all angles $\phi$.

The strong variation is also seen for the other molecules *n*-butane and 2-methyl-pentane, cf. Figure 2. Here, the descriptor value for a non-rotated molecular structure ($\phi=0$ for all coordinate axes) is compared to the maximum and minimum values obtained during full rotations (0° - 360°) around all three coordinate axes. At an angle of 0°, the descriptor values allow a distinction not only of different compounds but also of the conformations of a compound. The variation of descriptor values due to rotation, however, leads to partly indistinguishable compounds and conformations.

A similar behavior is found for descriptor RDF010u, which only depends on the interatomic distances of a molecule [8], cf. Figure 3. DragonX rounds atom coordinates after the 4[th] digit. Hence, after a rotation according to eq. (5) is performed, the novel atom coordinates are again rounded and thus a small numerical error is introduced which also affects the intermolecular distances. However, while such numerical errors should be of a similar magnitude, Figure 3 shows a periodic deviation from the descriptor values at $\phi = 0°$, which also differs for rotations around the 3 coordinate axes. This clearly indicates that the inaccuracy introduced by coordinate rotation is not only due to rounding errors, but that there has to be another reason which is still unknown. However, as Figure 4 shows, the relative deviation caused by rotation is much smaller than for DISPv, which is due to the larger absolute descriptor values. Although the single conformations thus remain distinguishable for RDF010u, the periodic descriptor variation along the rotation angle $\phi$ represent an unphysical behavior which casts the reliability of the RDF010u descriptor into doubt.

The DISPv and RDF010u descriptors have been chosen only as representatives, and further 3D descriptors evaluated by DragonX show similar behavior. But regardless of the number of descriptors exhibiting the observed behavior, any molecular descriptor which is systematically affected by the coordinate system used to represent a molecule has to be considered unphysical and should not be used within a QSPR model, since its true information content is unknown. Therefore, it does not allow to derive a physical relation between macroscopically observed properties and the molecular characteristics which are represented by the molecular descriptor. While it is currently not clear whether the descriptor variations are due to insufficient descriptor definition or due to implementation issues within DragonX, it is highly encouraged to first resolve such issues before employing 3D descriptors in QSPR models.

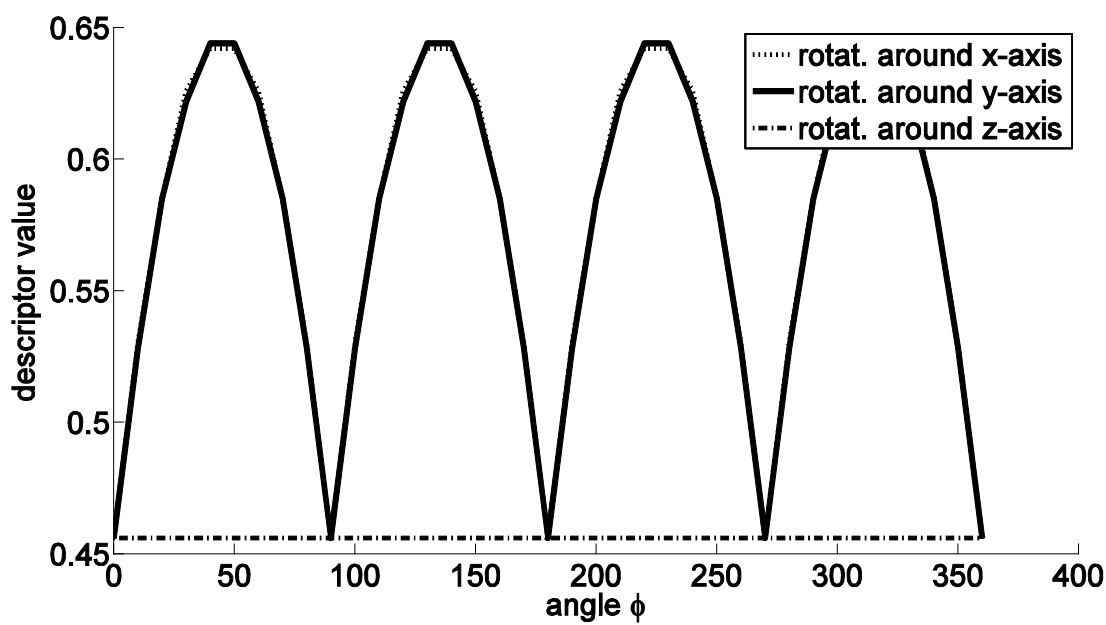

**Figure 1.** Evaluation of the molecular descriptor DISPv for *n*-propane as a function of the rotation angle around the *x*- *y*- and *z*-axis. Evaluations have been performed between 0° and 360° for all three axes.

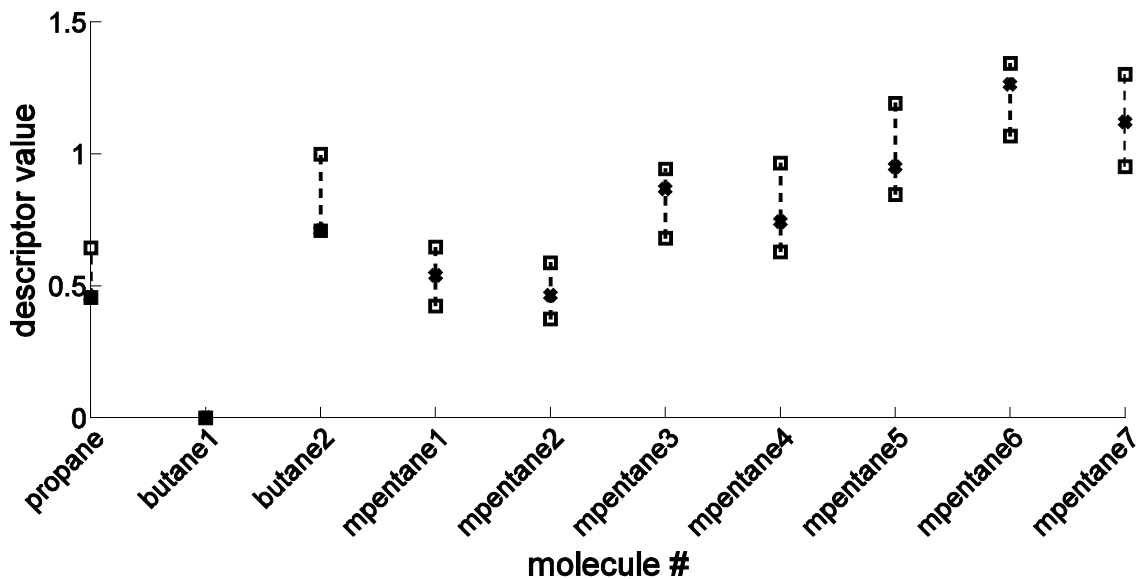

**Figure 2.** Descriptor DISPv of all conformations of *n*-propane, *n*-butane and 2-methylpentane. The black cross shows the descriptor evaluated at the original coordinates (i.e. 0° rotation in all three axes), while the intervals show the maximum descriptor variation when rotating the original structure by 360° in all three coordinate axes.

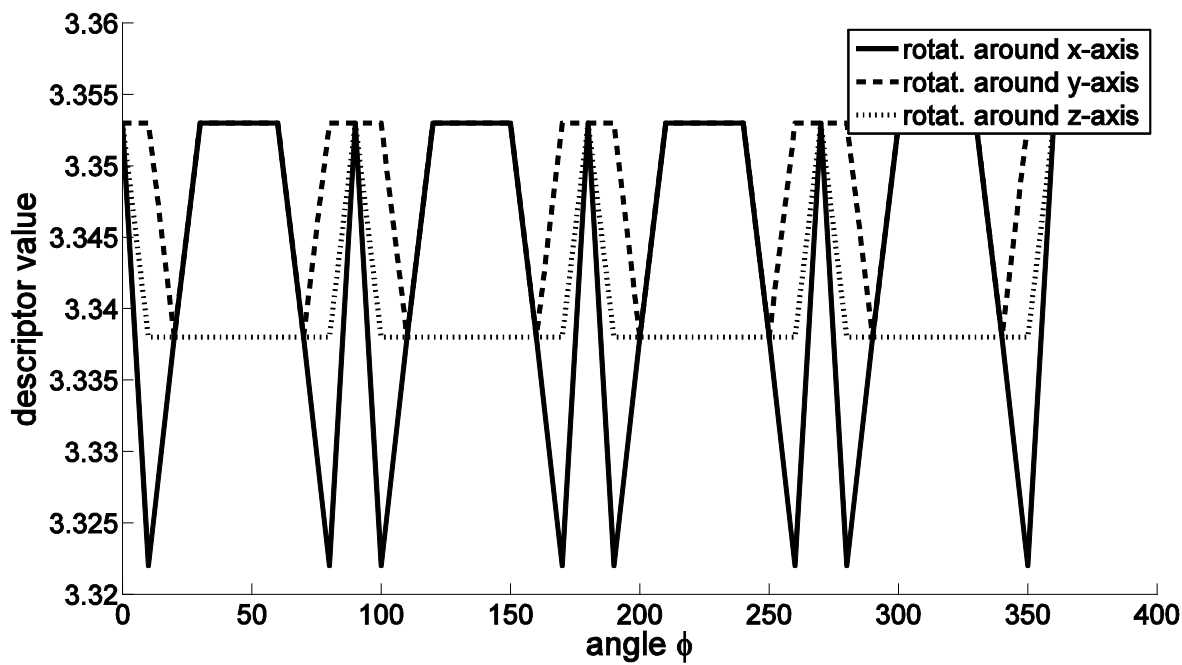

**Figure 3.** Evaluation of the molecular descriptor RDF010u as a function of the rotation angle around the x- y- and z-axis. Evaluations have been performed between 0° and 360° for all three axes.

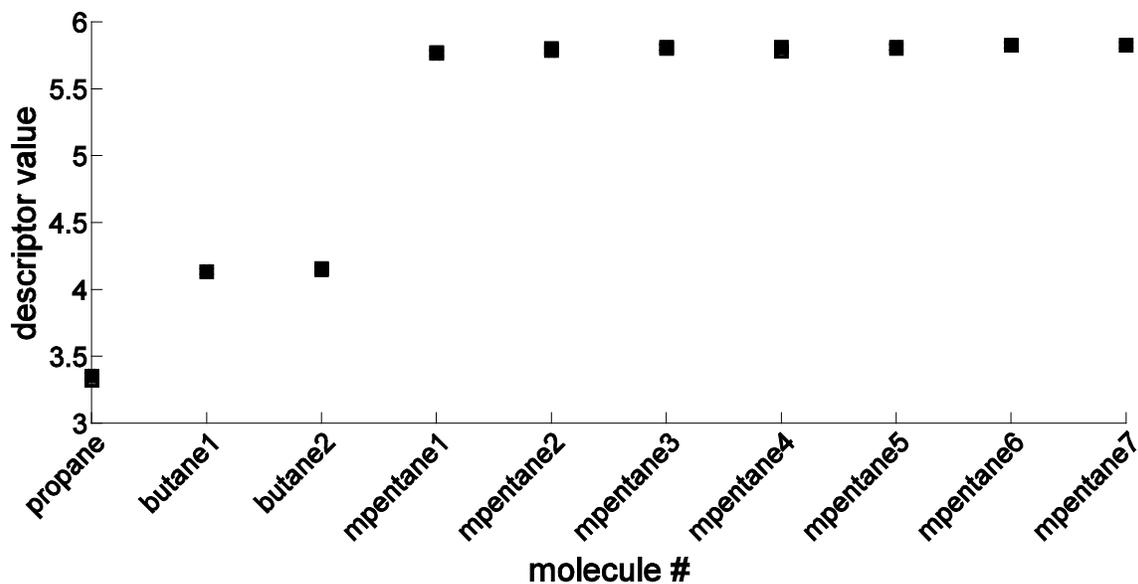

**Figure 4.** Descriptor RDF010u of all conformations of *n*-propane, *n*-butane and 2-methylpentane. The cross shows the value evaluated the original coordinates (i.e. $\phi = 0°$ rotation), while the intervals show the maximum descriptor variation when rotating the original structure by $\phi = 360°$ in all three axes.

## Summary and conclusions

The variation of several molecular descriptors which depend on the spatial molecular orientation has been investigated with respect to rotation of the underlying coordinate system used to describe the atomic coordinates. To this end, a systematic conformer search has been combined with geometry optimization on a B3LYP/TZVP level for the compounds *n*-butane, *n*-propane and 2-methylpentane. The converged geometries of the obtained conformations have been rotated around all 3 coordinate axes, and during each rotation 3D molecular descriptors have been evaluated using the widely employed descriptor calculation package DragonX.

The results clearly indicate that rotating the coordinate system which is used to describe the molecular structure severely affects the descriptor values in an unphysical way. Since any rotation does not affect the molecule itself, the obtained descriptor values should be invariant to rotation. Until now, it is not clear whether the descriptor variations result from insufficiently well-defined molecular descriptors or if they are due to the implementation of the descriptors within DragonX. But regardless of the reason for the observed behavior, any 3D descriptor exhibiting sensitivity with respect to the coordinate system used for molecular representation yields highly questionable molecular information which can hardly be used for reliable QSPR modeling.

## Acknowledgement

This work was performed as part of the Cluster of Excellence "Tailor-Made Fuels from Biomass", which is funded by the Excellence Initiative by the German federal and state governments to promote science and research at German universities.

## References

[1] M. Hechinger, K. Leonhard, W. Marquardt: What is Wrong with QSPR Models Based on 3D Descriptors?. *Journal of Chemical Information and Modeling*, **2012**, 52(8), 1984–1993.

[2] Dewar, M.; Zoebisch, E.; Healy, F.; Stewart, J. *J. Am. Chem. Soc.* **1984**, 107, 3902-3909.

[3] Spartan10, version 1.1.0, Wavefunction, Inc., Irvine CA, **2011**.

[4] Gaussian 09, Revision A.02. Frisch, M. J. et al., Gaussian, Inc., Wallingford CT, **2010**.

[5] DragonX for Windows, version 1.4, Talete srl., Milan Italy, **2009**.

[6] Chirlian, L. E.; Francl, M. M. *J. Comput. Chem.* **1987**, 8, 894-905.

[7] *http://www.tripos.com*, accessed on Oct. 30[th], 2012.

[8] Todeschini, R.; Consonni, V. Molecular descriptors for chemical informatics; Mannhold, R.; Kubinyi, H., Folkers, G., Eds.; Wiley: Weinheim, **2000**.